\documentclass[10pt]{elsart}

\usepackage[]{graphicx}
\usepackage{amssymb}
\usepackage[]{appendix}

\begin{document}
\begin{center}

\begin{frontmatter}

\title{Transverse intrinsic localized modes in monatomic chain and in graphene}

\author[Tartu]{V. Hizhnyakov},
\author[Tallinn]{M. Klopov},
\author[Tartu]{A. Shelkan\corauthref{cor}}
\corauth[cor]{Corresponding author.} \ead  {shell@ut.ee}

\address[Tartu]{Institute of Physics, University of Tartu, Ravila 14c,
50411 Tartu, Estonia}
\address[Tallinn]{Department of Physics, Faculty of Science,
Tallinn University of Technology, Ehitajate 5, 19086 Tallinn, Estonia}

\begin{abstract}

In this paper an analytical and numerical study of anharmonic vibrations of
monatomic chain and graphene in transverse (perpendicular) with respect
to the chain/plane direction is presented.
Due to the lack of odd anharmonicities and presence of hard
quartic anharmonicity for displacements in this direction, there may exist
localized anharmonic transverse modes with the frequencies above the
spectrum of the corresponding phonons. Although these frequencies are in
resonance with longitudinal (chain) or in-plane (graphene) phonons, the
modes can decay only due to a weak anharmonic process. Therefore
the lifetime of these vibrations may be very long. E.g. in the chain,
according to our theoretical and numerical calculations it may
exceed 10$^{\mathrm{10}}$ periods. We call these vibrations
as transverse intrinsic localized modes.
\end{abstract}


\end{frontmatter}
\end{center}

\section{Introduction}

It is a well-known fact that point defects in crystals may cause an
appearance of spatially localized vibrations, called local modes (see, e.g.
\cite{Maradudin71}). The frequency of such modes lies outside the phonon spectrum,
which prevents the spreading of the vibrations to the bulk. Besides, resonant or
pseudolocal modes with the frequencies inside the phonon spectrum
may also exist. They appear when the frequency of the leading vibration
associated with the defect gets to a region of small density of states (DOS)
of phonons. Unlike local modes, which are stable in harmonic approximation,
these modes can live only a finite time due to the emission of resonant
phonons. However, because of the small DOS of resonant phonons the
interaction of the latter with pseudolocal mode is weak, due to what the lifetime
of the mode becomes long. A well-known case of long-living pseudolocal modes
in 3D lattices is the one by the defect of large mass \cite{Maradudin71}.

Spatially localized modes may exist also in perfect nonlinear lattices. Such
modes in chains with cubic and quartic anharmonicity were first described by
A. M. Kosevich and A. S. Kovalev \cite{KosKov}. They have found that for sufficiently
strong quartic anharmonicity there exist vibrational states localized in the
space and periodic in time; the frequency of these vibrations exceeds the
maximum frequency of phonons. The authors \cite{KosKov} restricted their consideration
with large-size vibrations and the frequency close to the top of the phonon
spectrum. Small size localized vibrations in anharmonic lattices were
introduced by Dolgov \cite{Dolgov} and then by A. J. Sievers and S. Takeno
\cite{Sievers88} and were called ``intrinsic localized modes'' (ILMs).

In the numerical studies of ILMs different two-body potentials
(Lennard-Jones, Born-Mayer-Coulomb, Toda, and Morse potentials and their
combinations) have been used (see, e.g. \cite{Page, Sievers95, Flach}).
All these potentials have
strong odd anharmonicities and show a strong softening with the increasing
of the vibrational amplitude. The ILMs found in these simulations always
drop down from optical band(s) into the phonon gap, if there is any. In this
connection see Ref. \cite{Kiselev}, where gap ILMs in NaI were calculated
without taking into account long-range interactions and Ref. \cite{Haas10},
where the calculations of ILMs in NaI were made taking these interactions into
account.

Usually in crystal lattices odd anharmonicities are strong and pair
potentials show a strong softening with the increase of vibrational
amplitude. Therefore, the dropping down of the frequency of ILMs from
optical bands is quite common in 3D lattices. Still, as it was shown in
Refs. \cite{Voulgarakis, Hizhnyakov15, Haas11, Hizhnyakov14},
in some crystals odd anharmonicities are reduced due to
multiparticle or covalent interactions. Examples have been given with
germanium \cite{Voulgarakis}, diamond \cite{Hizhnyakov15}, metallic Ni, Nb
\cite{Haas11}, iron \cite{Hizhnyakov14} and copper  \cite{Hizhnyakov15};
in all these crystals ILMs with frequencies above the top of the
phonon band were found in numerical simulations.

However, there are systems in which the odd anharmonicities disappear due to symmetry
arguments; at the same time, the quartic anharmonicity is non-zero and it is hard
(positive). The examples of such systems are given by linear atomic chains and planar
atomic structures (e.g. graphene); in these systems odd anharmonicities disappear
for vibrations in the transverse (out-of-chain and out-of-plane) direction.
Therefore, one can expect that in these systems there can exist transverse anharmonic
local modes with the frequencies above the maximum frequency of the corresponding phonons.
The latter frequencies are usually smaller than the maximum frequency of
longitudinal/in-plane phonons. These modes fall in resonance with these phonons
and can decay. However, unlike pseudolocal modes and like local modes the interaction
causing the decay of these modes is anharmonic. Therefore we call them as transverse ILMs  (TILMs).
The anharmonic interaction of TILMs with small vibrational amplitudes
of atoms is very weak. Hence, one can expect that the lifetime of such TILM may be
rather long. Below we will consider TILMs in a monatomic chain and in graphene,
both analytically and numerically. Our considerations confirm the aforesaid expectation.

\section{Anharmonic chain}
First we consider the anharmonic monatomic chain and examine the vibrations
of its atoms in transverse ($y)$ directions. We suppose that the potential
energy $U$ of the chain is given by the sum of pair potentials $V(R_{n,n'})$, 
where $R_{n,n'}$ is the distance between the atoms $n$ and $n'$.
The latter potentials can be expanded into the series of atomic displacements.
Denoting $\ d=(n-n')\ a$, $\ x=x_{n}-x_{n'} $, $\ y=y_{n} -y_{n'}$,
where $a$ is the atom spacing, $x_{n} $ and $y_{n} $ are the longitudinal
and transverse displacements of the atom number $n$ from its equilibrium
position, we get $R \equiv R_{n,n'}=\sqrt{(d+x)^2+y^{2}}$.
As y appear as $y^2$, any power expansion of R will have only even powers of $y$.  
The same holds for $U$. This means that $U$ indeed has no odd anharmonic terms. This is 
a consequence of the symmetry of the chain with respect to the change of the sign of $y$.

Note one more property of the chain: the term in the expansion of $R$, quadratic
with respect to $y$, has the same numerical factor as the term linear with
respect to $x$. The same holds also for an arbitrary power of $R$. In the
equilibrium state all linear terms with respect to the coordinates $x$ in
the potential energy $U$ are cancelled. Therefore, all quadratic terms with
respect to $y$ are also cancelled, i.e the frequencies of transverse vibrations in the 
pair potential approximation tend to zero. As a result long-range fluctuations can be 
created with little energy cost and since they increase the entropy they are favored. 
This leads to the instability of the chain with respect to small transverse distortions 
(see in this connection the Mermin-Wagner theorem \cite{Mermin}).

To get the chain stable one needs to stretch it \cite{Cadet}. In this case
the atom spacing $a$ is replaced by $a+s$, where $s$ is stretching. Then
the terms in $U$ linear with respect to $x$ and $y^{2}$ are not cancelled any more.
Therefore, the elastic springs for transverse vibrations are also nonzero
and positive. This results in the appearance of transverse phonons with finite,
although small for small stretching maximum frequency $\, \omega_{tm} $ . As these
phonons do not have any cubic anharmonicity, but have nonzero positive
quartic anharmonicity, low-frequency TILMs with the frequency above the
spectrum of transverse phonons should exist here.

Let us consider the TILM in a monatomic chain with the Morse pair potential
\begin{equation}
\label{eq1}
V=D( {1-e^{\alpha (a-r)}} )^{2}.
\end{equation}
Here $D$ is the energy of
dissociation, $\alpha $ is the parameter. We are using dimensionless
coordinates with the units corresponding to $a=1$ and the value $\alpha =4$
of the Morse pair potentials of atoms in monatomic metals. For this
potential only nearest-neighbor interactions are essential and only these
potentials will be taken into account here. We also take for the mass units
the mass of the atoms of the chain $(M=1)$. The dissociation energy is
chosen so that the unit frequency will correspond to the maximum frequency
of longitudinal phonons. In this case the potential energy of the stretched lattice 
is the sum of the following pair potentials (up to a constant term):

\begin{equation}
\label{eq2}
V=\Big( 1-e^{-4(r-1)} \Big)^{2} \bigg/ 128
- x \left( 1-e^{-4s} \right) e^{-4s} \Big/ 16,
\end{equation}
where $r=\sqrt {(x+1+s)^{2}+y^{2}} $. The last term in Eq. (\ref{eq2}) accounts 
for the effect of the stretching force of the chain in $x$ direction - 
it changes the equilibrium distance of the atoms 
from $r=1$ to $r=1+s$. Let us expand the potential into the series of $x$ and $y$ 
and take into account up to the second-order terms with respect to $x$ 
and forth-order terms with respect to $y$. We get (up to a constant)

\begin{equation}
\label{eq3}
V(x,y)\approx \frac{\nu_{1}}{8} x^{2} + \frac{\nu_{2}}{8} sy^{2}
+ \frac{\nu_{3}}{8} xy^{2} + \frac{\nu_{4}}{32} y^{4},
\end{equation}
where $\nu_{i}$ are dependent on stretching  $s$ parameters. In the
small $\,s\,$ limit $\ \nu_{i} \approx 1$. If $\,s=0.05\ $ then
$\,\nu_{1} \approx 0.522$, $\,\nu_{2} \approx 0.707$,
$\,\nu_{3} \approx 0.463$, $\,\nu_{4} \approx 0.441$.

The pair potential of the longitudinal vibrations alone is given by the first term in the 
right-hand side of Eq. (\ref{eq3}). Vibrational frequencies of corresponding phonons equal 
\cite{Maradudin71}

\begin{equation}
\label{eq4}
\omega_{k} =\sqrt {\nu_{1} \left( {1-\cos \left( k \right)} \right) / 2}.
\end{equation}
The maximum frequency of longitudinal phonons corresponds to $k=-\pi$ and equals 
$\omega_{lm}=\sqrt{\nu_1}$. The transverse vibrations alone are described by the 
pair potential

\begin{equation}
\label{eq5}
V(0,y) = \frac{\nu_{2} s}{8}\, y^{2} + \frac{\nu_{4}}{32}\,y^{4}\,.
\end{equation}
In harmonic approximation ($\nu_4 = 0$) the frequencies of corresponding phonons 
are given by Eq. (\ref{eq4}) with $\nu_2 s$ instead on $\nu_1$. The positive quartic 
anharmonicity in Eq. (\ref{eq5}) leads to appearance of the anharmonic modes  
\cite{KosKov}  (called here as TILMs) with the frequencies
$\omega_{0} =\omega_{tm} \sqrt {1+\varepsilon^{2} \big/ 4} $
above the maximum frequency of transverse phonons
$\ \omega_{tm} =\sqrt {\nu_{2} s} \ $ and with the displacements

\begin{equation}
\label{eq6}
y_{n} (t) \approx (-1)^{n} A_{0} \,\cosh^{-1}(\varepsilon n)\cos (\omega_{0} t) \ .
\end{equation}
Here $A_{0} $ is the amplitude of the central atom,

\begin{equation}
\label{eq7}
\varepsilon = \sqrt {3\nu_{4} } \, A_{0} \big/ \omega_{tm}
\end{equation}
is the reversed size of the TILM (we use the discrete analog of the derived 
in Ref. \cite{KosKov} equation (47) for the difference in the displacements
of two neighbouring atoms $\chi (x)$ ). 
These modes interact with longitudinal phonons and, therefore, they decay. To describe
this decay we consider the longitudinal vibrations of atoms in the presence of the TILM.
Taking into account Eq. (\ref{eq6}) we replace $y$ by $y(t)$ in Eq. (\ref{eq3})
and get the following pair potential for this motion:

\begin{equation}
\label{eq8}
V(x,t) \approx \frac{\nu_{1} }{8} x^{2} + \frac{\nu_{3}}{16} x y^{2}(0)
+ \frac{\nu_{3}}{16} xy^{2}(0) \cos (2\omega_{0} t).
\end{equation}
The first term in the right-hand side of this equation gives the potential energy of the 
longitudinal vibrations alone in harmonic approximation. The second and the third terms 
describe the anharmonic interaction of these and transverse vibrations. 
At that the second term stands for a small local compression, while the third term
describes the force with the frequency $2\omega_{0}$ periodically changing in time.
For all longitudinal phonons, except those with the resonant frequency 
$\omega_{k} =2\omega_{0} $, this force causes forced vibrations of atoms with 
the frequency $2\omega_{0} $.
The resonant term causes the increase of the energy of phonons in time. 
From energy conservation law it follows that this energy comes from the TILM, 
i.e. the TILM decays.

\section{Decay rate of transverse ILM in chain}
To find the rate of decay we are considering the equation of motion of the
longitudinal phonon with the coordinate
\begin{equation}
\label{eq9}
x_{k} =N^{-1/2} \sum\nolimits_n {x_{n} \sin (kn)},
\end{equation}
where $N \gg 1$ is the number of atoms in the chain,
$k=2 \pi m \big/ N$ is the wave number of the phonon, $m=0,\,\pm 1,\,\pm 2,\,...$
(we use periodic boundary conditions), $x_n$ is the displacement of the atom $n$ from its 
equilibrium position. In the case of pair potential (\ref{eq8}) this equation reads

\begin{equation}
\label{eq10}
\ddot{{x}}_{k}  \cong  - \omega_{k}^{2} x_{k} +\frac{\nu_{3}}{16 N}
\sum\limits_{n=-\infty}^\infty {\sin(kn) \Big( {(y_{0,n+1} -y_{0,n} )^{2}
-(y_{0,n} -y_{0,n-1} )^{2}} \Big) \cos(2\omega_{0} t)},
\end{equation}

where $\omega_{k}$
is the frequency of the longitudinal phonon $k$ given by Eq. (\ref{eq4}),  
$y_{0,n} \equiv y_{n} (0)$ is the initial amplitude of the transverse  ILM on atom $n$ 
given by Eq. (\ref{eq6}). The first term in the right-hand side of Eq. (\ref{eq10}) 
stands for the harmonic force stemming from the first term in the right-hand side 
of Eq. (\ref{eq8}) and the second term accounts for the anharmonic force stemming 
from the third term in the right-hand side of this equation 
(a small compression of the chain stemming from the second term in the right-hand side 
of Eq. (\ref{eq8}) is neglected). The transverse amplitude $ | y_{0,n} | $
slowly changes with $n$. Therefore, $ \ y_{0,n+1} - y_{0,n} \cong 2 {y}_{0,n} $ and

\begin{equation}
\label{eq11}
(y_{0,n+1} - y_{0,n} ) ^ 2 - (y_{0,n} - y_{0,n-1} ) ^ 2
\approx 4 \, {\partial y_{0,n}^{2} } \Big/ {\partial n} \, .
\end{equation}
Inserting Eq. (\ref{eq11}) into Eq. (\ref{eq10}) and taking into account Eq. (\ref{eq6}), we get

\begin{equation}
\label{eq12}
\ddot{{x}}_{k} +\omega_{k}^{2} \, {x}_{k} \cong C_{k} N^{-1/2} \cos (2\omega_0 \,t),
\end{equation}
where
\begin{equation}
\label{eq13}
C_{k} \cong \frac{\nu_{3} }{4} A_{0}^{2}
\sum \limits_{n=-\infty}^\infty \sin (kn) \frac{\partial }{\partial n}
\left( \frac{1}{\cosh^{2}(\varepsilon n)} \right)  \cong
- \varepsilon \nu_{3} A_{0}^2 \sum\limits_{n=0}^\infty
\sin(kn) \frac{\sinh (\varepsilon n)}{\cosh^{3}(\varepsilon n)} \ .
\end{equation}
By using the Green's function of harmonic oscillator $G_{k} (t)=\omega_{k}^{-1}
\sin (\omega_{k} t)$, the solution of Eq. (\ref{eq12}) can be presented in
the following form:

\begin{equation}
\label{eq14}
x_{k} (t)=x_{0,k} \cos (\omega_{k} t)+{x}'_{k} (t),
\end{equation}
where

\begin{equation}
\label{eq15}
{x}'_{k} (t)=\frac{C_{k} }{\sqrt N \omega_{k} } \int\limits_0^t {\sin
\left({\omega_{k} (t-t')} \right)} \cos (2\,\omega_{0} \,t')\,dt'\,.
\end{equation}
The first  term in Eq. (\ref{eq14}) ($\propto x_{0,k} )$ stands for free oscillations, 
while the second term (${x}'_{k} (t))$ describes the excitation of vibrations by the
transverse mode. The energy of the excited vibrations is given by the sum of the terms 
$\ \omega_{k}^{2} \, {x'}_{k}^{2} \ $ averaged over period and summed over all phonons. 
This gives

\begin{equation}
\label{eq16}
E(t) = \frac{1}{N} \sum \limits_k C_{k}^{2}
\Big\langle \! \int\limits_0^t \! dt' \!
\sin \left(\omega_{k} (t-t') \right) \cos \left(2 \omega_{0} t' \right)
\int\limits_0^t \! dt'' \!
\sin \left(\omega_{k} (t-t'') \right) \cos \left(2 \omega_{0} t'' \right)
\! \Big\rangle
\end{equation}
Let us take into account the relations:

\[
\sin (\omega_{k} (t-t') ) \sin (\omega_{k} (t-t'') )
= (\cos (\omega_{k} (t'-t'') ) -
\cos ((\omega_{k} (2t-t'-t'') ))\big/ 2 \, ,  \]

\[
\cos (2 \omega_{0} t') \cos (2 \omega_{0} t'' )
= (\cos (2 \omega_{0} (t'-t'') ) +
\cos (2 \omega_{0} (t'+t''))) \big/ 2 \, .
\]

In the case under consideration frequencies of phonons form a continuous spectrum. 
In this case, due to the summation (integration) over $k$ the terms in Eq. (\ref{eq16}) 
$\propto \cos(\omega_{k}(t'-t''))$ and $\propto \cos((\omega_{k}(2t-t'-t''))$ 
essentially differ from zero only for finite values of the arguments $t'-t''$ 
and $2t- t'-t''$ (of the order of unity). At that the term 
$\propto \cos((\omega_{k}(2t-t'-t'')))$ essentially differs from zero only 
for large $t'$,  $t'' \sim t$ when $\cos (2\omega_{0}t')$ and $\cos (2\omega_{0}t'')$ 
average the contrubution of this term out. Neglecting  now unessential for 
$t',t'' \sim t>>\omega_0^{-1}$ term $\propto \cos (2 \omega_{0} (t'+t'') ))$  
we come to conclusion that at large $t$ only the term 
$\propto \cos (2 \omega_{0} (t'-t''))$ can give a remarkable contribution 
to the energy $E(t$). Finally we take into account that at large time $t$ in the  factor

$$
\cos (\omega_{k} (t'-t'')) \cos (2 \omega_{0} (t'-t'')) \!= \! (\cos((\omega_{k} -2 \omega_0) (t'-t''))+\cos((\omega_{k} +2 \omega_0) (t'-t'')))\big/2
$$

only the term $\cos ((\omega_{0}-2 \omega_{0}) (t'-t''))/2$
is essential.
This gives

\begin{equation}
\label{eq17}
E(t) \cong  \frac{1}{8 N} \sum \limits_k C_{k}^{2}
\int\limits_0^t dt' \int\limits_0^t dt''
\cos \left((\omega_{k} -2 \omega_0) (t'-t'') \right)
\end{equation}
For small $ | t'-t'' | $ being essential here one can replace the integral
$\int\limits_0^t \int\limits_0^t dt' dt'' =
\int\limits_0^t dt_1 \int\limits_{-t'}^{t'} d\tau $
by the integral $\int\limits_0^t dt_1 \int\limits_{-\infty}^\infty d\tau $,
where $t_1 = (t'+t'') / 2$, $\tau = t'-t''$.
As a result, in the $t \to \infty $ limit the rate of increasing
of the energy of longitudinal phonons, followed by a subsequent decrease
of the energy of the TILM under consideration, equals

\begin{equation}
\label{eq18}
dE(t) /dt = \left( \pi / 4 N \right) \sum\limits_k C_{k}^{2}
\delta (\omega_{k} -2\omega_{0} ) ,
\end{equation}
Replacing here the sum by the integral and taking into account the relations
$dm / dk =N / 2 \pi \,$ and $ \,dk / d\omega_{k} = 4 \omega_{k} / \sin k$ one finds

\begin{equation}
\label{eq19}
dE / dt \approx \left( \omega_{0} \nu_{3}^{2} A_{0}^{4} \Big/ \sin k_0 \right)
\Phi^{2} \left( \Omega \right),
\end{equation}
where $\omega_{0} =\omega_{tm} \sqrt {1+\varepsilon^{2} \big/ 4} \, $,
$k_{0} =\arccos \left(1-8\omega_{0}^{2} / \nu_{1} \right)$,
$\Omega = k_0 / \varepsilon $,
$ \varepsilon = A_{0} \sqrt{3\nu_{4}} \big/ \omega_{tm} $,

\begin{equation}
\label{eq20}
\Phi (\Omega )=\int\limits_0^\infty {\frac{\sin \left( {\Omega x}
\right)\sinh(x)}{\cosh^{3}(x)}dx} .
\end{equation}

The decay constant $\Gamma $ of the TILM is determined by the equation
$\Gamma =E^{-1}dE /dt$. Taking into account the relation
$E= \omega_{0}^{2} \sum\nolimits_n y_{n}^{2} (0) /2
= \omega_{0}^{2} A_{0}^{2} \big/ \varepsilon $, we get
\begin{equation}
\label{eq17}
\Gamma =\frac{\varepsilon \nu_{3}^{2} A_{0}^{2} }
{\omega_{0} \sin k_0} \Phi^{2} \left( \Omega \right).
\end{equation}
$\Gamma $ decreases with decreasing of the amplitude of the central atom
$A_{0} $ of the TILM.
To get an estimation we consider the TILM with the amplitude $\, A_{0} =0.03 \,$
in the 5{\%} stretched chain $\,(s=0.05) $. In this chain
$\, \omega_{tm} \approx 0.1880$, $\, \varepsilon \approx 0.183$,
$\, \omega_{0} \approx 0.1888$, which gives  $\, \Omega \approx 6 \, $ and
$\, \Phi^{2}\left( \Omega \right)\approx 2\cdot 10^{-5}$. As a result we get
$\, \Gamma \approx 5\cdot 10^{-9}$. This corresponds to a very long lifetime.

\section{Transverse ILM in the chain; numerical study}
\label{subsec:transverse}

In our numerical study of transverse ILMs we considered a 5{\%} stretched
chain with
40 000 atoms and with fixed ends. The small compression of the entire chain
due to TILM was taken into account. The potential energy was described by
the interactions of the nearest neighbors with the Morse pair potential
given by Eq. (\ref{eq1}). We have found that TILMs indeed exist in this
chain and they have a long lifetime (see Figs. 1-3; one period of the TILM
with the frequency $\omega_{0} \approx 0.1888$ corresponds to 33 time units).
E.g. in our numerical simulations we have found that the TILM with $A_{0}
=0.03$ and $s=0.05$ has the frequency $\omega_{0} \approx 0.1883$ and the reversed
size parameter $\varepsilon \approx 0.113$. It decays very slowly: its amplitude
diminishes less than $2\cdot 10^{-8}$ for 1000 periods of vibrations. This
corresponds to $\Gamma < 2\cdot 10^{-11}$, which indeed gives very long
lifetime, in agreement with the conclusion of the theory.

\vspace{1.5cm}

\begin{figure}[h]
\begin{center}
\includegraphics[angle=-90,width=.79\textwidth]{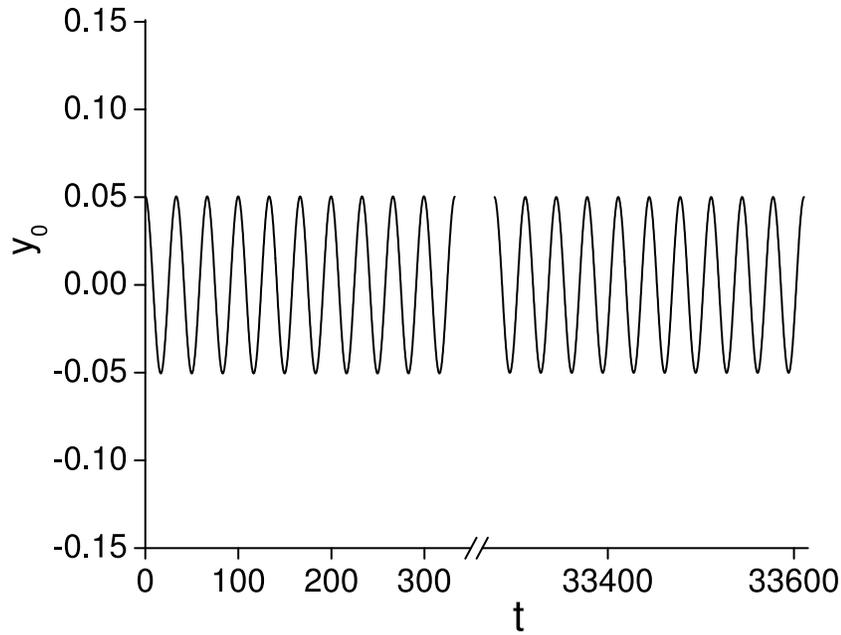}\hspace*{0em}
\end{center}
\caption {Time dependence of the transverse shifts (in the units of atom spacing) 
of the central atom of a transverse IPLM in 5{\%} stretched monatomic chain; 
the initial shift of the central atom is 0.05. 
First 10 periods and 10 periods after 1000 periods of vibration are presented.}
\vspace{0.20cm}
\end{figure}

The TILMs with the amplitudes $A_{0} > 0.05$ decay faster;
the rate of their decay rapidly increases with the increasing of the
amplitude. This is also in agreement with the theory.

\vspace{0.70cm}

\begin{figure}[h]
\begin{center}
\includegraphics[angle=-90,width=.79\textwidth]{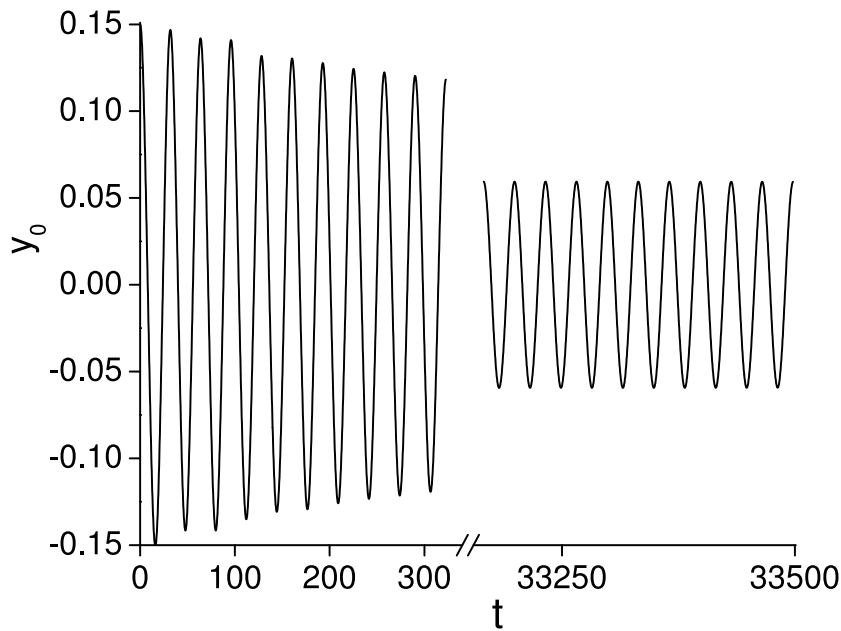}\hspace*{0em}
\end{center}
\caption {The same as in Fig. 1, but the initial shift of the central atom
of a transverse ILM is 0.15.}
\end{figure}

\begin{figure}[h]
\begin{center}
\includegraphics[angle=-90,width=.99\textwidth]{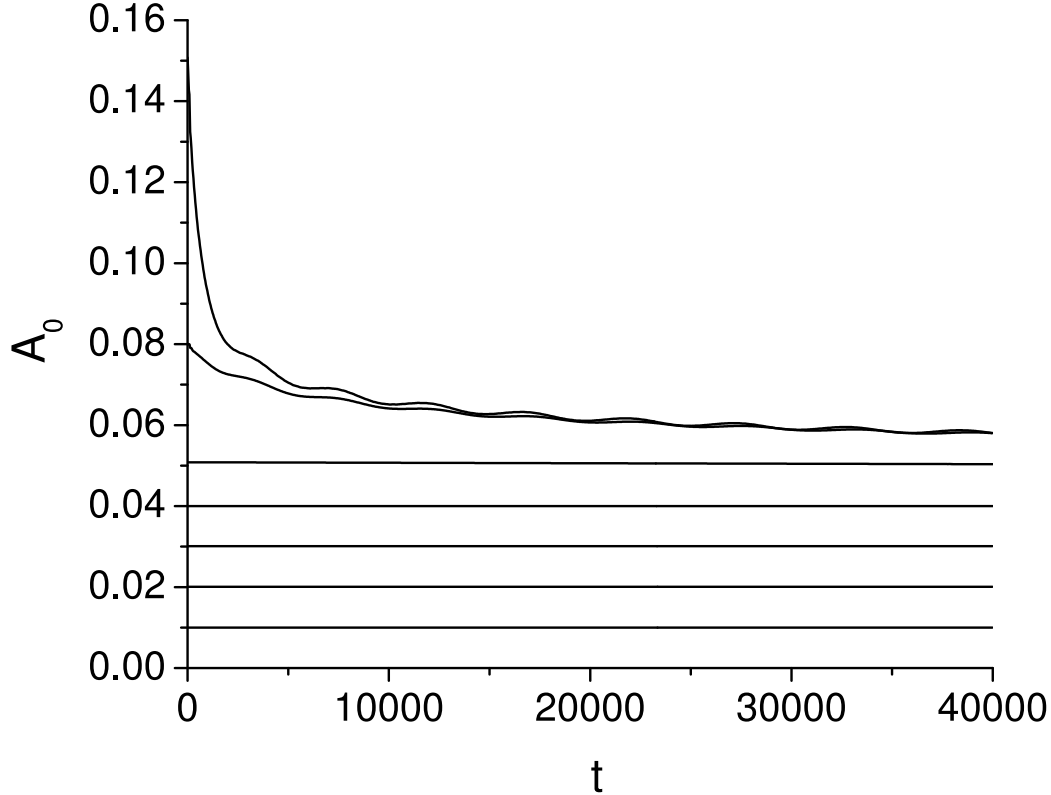}\hspace*{0em}
\end{center}
\caption {Time dependence of the amplitudes of the central atom of transverse ILMs 
in 5{\%} stretched monatomic chain for different initial shifts of the central atom.}
\vspace{0.5cm}
\end{figure}

Concerning the difference of the above-presented numerically-found values of
the parameters $\varepsilon $ and $\omega_{0} $ from their theoretical
values, we have found that these differences come from the neglecting of the
local compression of the chain in the theory. This follows from the
numerical simulations of TILMs in the chain rigid in $x$ direction (with $x$
coordinates of atoms being fixed at their values without TILM), performed by
us. We have found the TILM in this chain with $\varepsilon \approx 0.183$ and
$\omega_{0} \approx 0.1888$ in full agreement with the above theoretically found
values of these parameters. This TILM does not decay, as it should according
to the theory presented above. An approximate account of the compression
mentioned in the theory, allowing one to explain the values $\varepsilon
\approx 0.113$ and $\omega_{0} \approx 0.1883$ found in simulations, is given in
Appendix. Note that if to use these values of $ \varepsilon $ and $\, \omega_{0}$
in   Eq. (\ref{eq17}) we get  $\, \Gamma \approx 2 \cdot 10^{-13}$.
This also corresponds to a very long lifetime.

\vspace{3cm}

\section{Graphene: out-of-plane vibrations}

The arguments presented in the beginning of Section 2 can readily be
extended to planar atomic lattices. To be specific, we are considering here
a graphene sheet. It consists of carbon atoms constituting the
honeycomb-type periodical structure. Every carbon atom in this structure is
connected with three neighboring atoms by chemical bonds formed by the
sp$_{\mathrm{2}}$ hybrid orbitales.

We suppose that in equilibrium positions the graphene is situated in the $(x,y)$
plane. The distances between an atom and three neighboring atoms equal
\begin{equation}
\label{eq18}
r_{n} =\sqrt {\left( {b_{n} a+x_{n} } \right)^{2}+\left( {c_{n} a+y_{n} }
\right)^{2}+z_{n}^{2} }
\end{equation}
($n=1,2,3)$, where $x_{n} $, $y_{n} $ and $z_{n} $ are the $x$-, $y$- and
$z$- components of the displacements of three neighboring atoms with respect
to the central atom, $b_{1} =0$, $ b_{2} ={\sqrt 3 } / 2$, $b_{3}
={-\sqrt 3 } / 2 $ , $c_{1} =-1$, $c_{2} =1 / 2$, $c_{3} =1 / 2$. In
the pair potential approximation the potential energy of the vibrations of
atoms depends on the distances $\, R_{n} =r_{n} -a \, $ (explicit
form of this energy in harmonic approximation, see e.g. in \cite{Adamyan}).
The expansion of these distances over the displacements $x_{n} $, $y_{n} $ and
$z_{n} $ depends on the powers
$R^{m}=\left(bx+cy+(x^{2}+y^{2}+z^{2}) / a \right)^{m}\,$
(here the subscript $n$ is omitted for simplicity).
In the equilibrium position the $\propto R$ terms are cancelled.
Therefore, the expansion of a pair potential over $z$ at $x=y=0$
starts with the positive quartic term $\propto z^{4}$,
i.e. the $\propto z^{2}$ terms are absent. Consequently, in the pair potential
approximation the 2D lattice is unstable with respect to small out-of-plane
distortions. However, in graphene the atomic interactions are determined by
covalent forces. These forces cause the stiffness of the planes with respect
to the transverse short-range displacements of atoms \cite{Adamyan}.
In accordance with the Mermin-Wagner theorem \cite{Mermin} the plane of graphene
remains unstable with respect to long-range transverse distortions, resulting
in the appearance of ripples \cite{Geim, Fasolino}.
A stretching of the graphene sheet removes the ripples.

The elementary cell of graphene includes two atoms. Therefore, there are two
branches of out-of-plane vibrations: acoustic and optic.
One can expect the existence of two types of out-of-plane TILMs in graphene
-- with the frequency above the top frequency of acoustic out-of-plane
phonons (the acoustic-like TILM) and with the frequency above the top
frequency of optic out-of-plane phonons (the optic-like TILM). The
acoustic-like out-of-plane TILMs can decay due to a relatively weak
anharmonic interaction with the in-plane phonons and also due to in general
stronger harmonic resonant interaction with out-of-plane optic phonons.
However, the optic-like out-of-plane TILMs can decay only due to a
relatively weak anharmonic interaction with in-plane phonons. Therefore, the
optic-type out-of-plane TILMs should have a longer lifetime.

This conclusion is in agreement with our numerical modelling of out-of-plane
TILMs in weakly-stretched (1.5 {\%}) graphene. We have used the AIREBO
potential and taken into account the cluster $246\times 210$ {\AA}
($100\times 50$ periods, 20 000 atoms) with periodical boundary conditions.
Initially we displaced out-of-plane six atoms of C$_{\mathrm{6}}$ ring
according to the optical mode. We found that out-of-plane optic TILMs indeed
exist; see Fig. 4 (where the time-dependences of the displacements of one of
the three equivalent atoms of the central C$_{\mathrm{6}}$ ring are
presented; the other three atoms in this ring vibrate out of phase with the
mentioned atom) and Fig. 5 (where in-plane and out-of-plane phonons DOS and
the spectrum of out-of-plane TILM for different initial amplitudes of carbon
atoms are given).

\vspace{0.5cm}

\begin{figure}[h]
\begin{center}
\includegraphics[angle=-90,width=.99\textwidth]{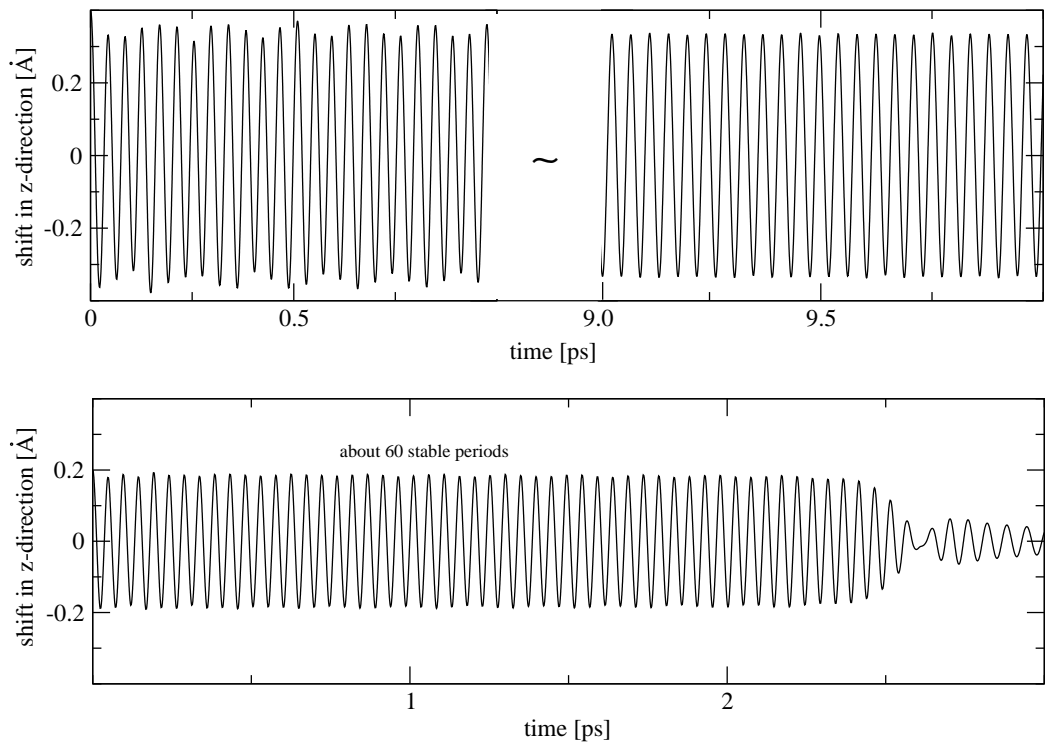}\hspace*{0em}
\end{center}
\caption {Optical-type out-of-plane transverse ILMs in graphene. The out-of-plane 
shifts (independent of time) of one of three equivalent atoms of the central
C$_{\mathrm{6}}$ ring is presented for initial amplitudes 0.4 {\AA} (upper)
and 0.2 {\AA} (bottom); the other three atoms in this ring vibrate out of phase
with the atom mentioned.}
\vspace{3.0cm}
\end{figure}

\vspace{3.0cm}

\begin{figure}[h]
\begin{center}
\includegraphics[angle=-0,width=.99\textwidth]{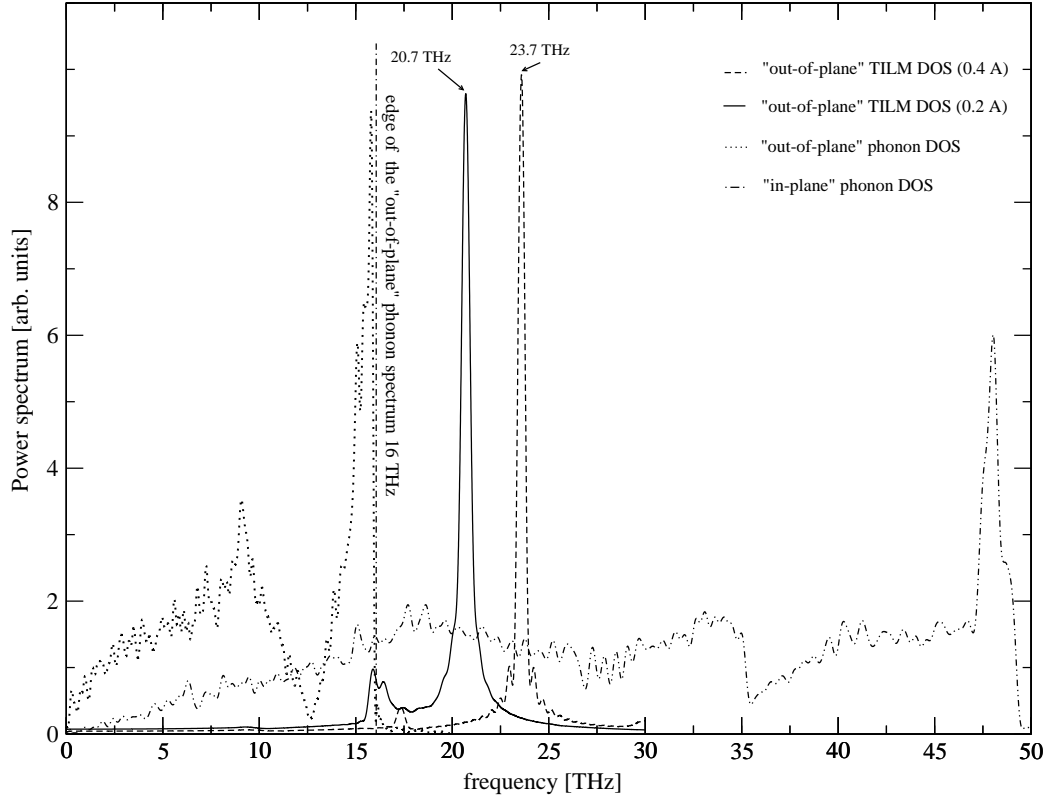}\hspace*{0em}
\end{center}
\caption {In-plane and out-of-plane phonon DOS for the AIREBO potential and the spectrum
of out-of-plane transverse ILM for different initial amplitudes of carbon atoms.}
\vspace{0.0cm}
\end{figure}

Note that in strongly-stretched graphene there can exist also in-plane
anharmonic localized vibrations (in-plane discrete breathers)
\cite{Khadeeva, Baimova}. It appears (see Ref. \cite{Khadeeva})
that strong uniaxial stretching (along the zigzag or armchair direction)
results in the opening of the gap in the middle of the phonon spectrum
of graphene. This makes possible the existence of in-plane soft ILMs
(discrete breathers) with the frequency in the gap of the phonon spectrum
\cite{Khadeeva, Baimova}. The authors \cite{Khadeeva, Baimova} have found
that these modes remain relatively stable also when, due to a large stretching,
their frequencies get into resonance with out-of-plane phonons.

Thus, the physical situation discussed in Refs. \cite{Khadeeva, Baimova}
is reversed to the one considered here: the anharmonic in-plane modes
with soft anharmonicity are in resonance with out-of-plane phonons,
while out-of-plain TILMs considered here are in resonance with in-plane
phonons. However, in both cases the lifetime of the modes is long.

\section{Discussion}
In this communication it was shown that due to symmetry arguments in linear
monatomic chains and in planar monatomic structures (e.g. graphene) odd
anharmonicities disappear for the motion of atoms normal with respect to the
chain/plane. However, even (quartic) anharmonicity exists and is hard
(positive). As a result, the anharmonic localized modes of transverse
(chain) or out-of-plane (graphene) vibrations (called TILMs) can exist in
these systems with frequencies above the transverse/out-of-plane phonons.
These vibrations have finite, although large lifetime: the slowly decay due to creation 
of longitudinal phonons. For the case of a monatomic chain a theory is being developed 
which allows one to calculate the decay rate of TILMs. The theory predicts that the
lifetime of the TILMs under consideration may be extremely long, longer than
10$^{\mathrm{10}}$ periods of vibration. The numerical modelling of these
TILMs fully confirms this result. The normal (out-of-plane) TILMs with the
frequency above the phonon spectrum of out-of-plane phonons were also found
for weakly-stretched graphene.

Here only immobile TILMs were considered. However, one can expect that these
excitations can move in the same way as ILMs can move in Cu, Ni and Fe
in the nearest neighbors crystallographic directions
\cite{Hizhnyakov15, Haas11, Hizhnyakov14}.

The symmetry argument presented here holds generally for linear atomic
chains and planar monolayer atomic structures. Therefore, one can expect the
existence of TILMs in any linear atomic chain and any planar
monolayer atomic structure. Besides, one can expect that an analogous
situation may exist in 3D lattices with dominating chain and layer
structural elements, e.g. in graphite and in mica. In such lattices odd anharmonicities
exist, but they may be essentially reduced for out-of-chain/plane
vibrations.

In this connection we would like to point to investigations of M. Russell et al. 
of muscovite mica irradiated by high energy particles (see, e.g. \cite{Russell}).  
The irradiation produces in mica very long black tracks parallel to the crystallographic 
plains and localized between the layers. These tracks are explained 
in Ref. \cite{Russell} by creation in the recoil process of discrete breathers 
which propagate on long distance along the crystallographic directions. 
The layered structure on atomic scale plays a crucial role in the observed phenomena. 
Although it is not yet clear whether these long-living vibrational excitations 
are transverse or longitudinal, the necessity of the layered structure on atomic scale 
to observe the phenomenon gives support to an assumption  that these excitation 
should contain remarkable transverse component.

Analogous TILMs may be expected to exist also for surface vibrations in
crystals with planar atomic surfaces. The change of the potential energy of
a surface atom for the shift in outer direction also does not have any odd
terms with respect to the shift. This leads to the decrease of the entire
odd-anharmonicity effects for out-of-plane surface vibrations, which may
result in the appearance of out-of-plane TILMs with frequencies above the
maximum frequency of transverse surface phonons and with a long lifetime.

It is known that next-to-next interactions in a chain can cause its
zigzag-type secondary structure \cite{Savin}. In such chains, e.g. in
polyethylene, the mobile ``solitons of tension'' may exist \cite{Manevitch}
(involving the longitudinal-type motion of atoms). Besides, in such chains
one should also expect the existence of transverse ILMs. Here two types of
transverse vibrations should be distinguished: parallel and perpendicular to
the plain of zigzag. The odd anharmonic terms for the transverse
displacements perpendicular to the plain mentioned are absent; however, for
transverse displacements parallel to this plain these are present. If the
zigzag angle is large, then the TILMs perpendicular to the plain of
zigzag should exist. However, if this angle is small, then the odd
anharmonicity for the displacements parallel to the plain of zigzag is
reduced. In such chains the existence of both TILMs is expected.
It is worthwhile to note that intrinsic localized modes (discrete breathers) are often 
used to explain the targeted energy transfer in such basic for living organism organic 
chains as DNA \cite{Bishop, Maniadis} and proteins \cite{Piazza}.

\textbf{Acknowledgements.} \ \  The research was supported by
Estonian research projects
SF0180013s07, IUT2-27 and by European Union through the European
Regional Development Fund (project 3.2.01.11-0029).

\begin{appendix}

\vspace{-0.5cm}
\section{The effect of self-compression on the shape of transverse ILM}

Let us consider the effect of longitudinal displacements of atoms, caused by
the transverse ILM, to its shape and frequency. From Eq. (\ref{eq8}) it follows
that the TILM causes the change (reduction) of the atomic distances by
$x_{c} =-\nu_{3} y^{2}(0) / 4\nu_{1}$ and slight vibrations of atoms
in $x$ direction with the frequency $2\omega_{0} $ (the forced vibrations).
Neglecting the small vibrations mentioned, we find that
on the average the $x$-dependent terms in Eq. (\ref{eq3}) give the following mean
(averaged over period of transverse vibrations) contribution to the
potential energy:

\begin{equation}
\label{eqa1}
\overline{V}_c =
\frac{\nu_{1}}{8} x_c^2 + \frac{\nu_{3}}{8} x_c \overline{y^2}
= \Big( \frac{\nu_{3}^{2}}{128\nu_{1}}
- \frac{\nu_{3}^{2}}{64 \nu_{1} } \Big) \, y^{4}(0)=
- \frac{\nu_{3}^{2}}{128 \nu_{1}} \, y^{4}(0)
\end{equation}

Let us find the nonlinear interaction $V_c$ of a transverse motion
with the mean value given by Eq. (\ref{eqa1}). We restrict ourselves
to the nonlinear interaction of $\propto y^{4}$ type. Taking into account
the relation $\,\overline {y^{4}} = 3y^{4}(0) / 8$, we get
$V_c = -\nu_{3}^{2} y^{4} / 48\nu_{1}$. We suppose that this potential
effectively accounts for the effect of displacements of atoms
in $x$ direction. Adding this term to $V(0,y)$ in Eq. (\ref{eq5}),
we get the following effective potential energy of the transverse vibrations
\begin{equation}
\label{eqa2}
V_{eff} (x_c, y) = V(0, y) + V_c =
\frac{\nu_{2} }{8}sy^{2}+\frac{\tilde{{\nu }}_{4}} {32} y^{4},
\end{equation}
where $\tilde{\nu }_{4} =\nu_{4} -2\nu_{3}^{2} / 3\nu_{1} $.
The TILM corresponding to this
potential energy is described by Eq. (\ref{eq6}) with the parameter
\begin{equation}
\label{eqa3}
\varepsilon =\sqrt {3\tilde{\nu }_{4}} \, A_{0} \Big/ {\omega_{tm}} \, .
\end{equation}
For the TILM with the amplitude $A_{0} =0.03$ of the central atom in 5{\%}
stretched chain $\tilde{{\nu }}_{4} \approx 0.167$. This gives
$\varepsilon \approx 0.113$ and $\omega_{0} \approx 0.1883$ in agreement with
numerically found values of $\varepsilon $ and $\omega_{0} $.
\end{appendix}

\end{document}